\documentclass[pra,twocolumn,prl,groupedaddress]{revtex4-1}

\usepackage{graphicx}
\usepackage{dcolumn}
\usepackage{bm}
\usepackage{epsfig}
\usepackage{amsmath}
\usepackage{amssymb}
\usepackage{color}
\usepackage{bbm}
\usepackage{braket}

\usepackage[colorlinks=true,citecolor=blue,linkcolor=blue,urlcolor=blue]{hyperref}

\begin{document}

\title{Light-matter interactions in synthetic magnetic fields: Landau-photon polaritons}

\author{Daniele De Bernardis$^1$, Ze-Pei Cian$^2$, Iacopo Carusotto$^3$, Mohammad Hafezi$^{2,4}$, Peter Rabl$^1$}
\affiliation{$^1$Vienna Center for Quantum Science and Technology, Atominstitut, TU Wien, 1040 Vienna, Austria}
\affiliation{$^2$ Joint  Quantum  Institute,  College  Park,  20742  MD,  USA}
\affiliation{$^3$ INO-CNR BEC Center and Dipartimento di Fisica, Universit\`a di Trento, I-38123 Povo, Italy}
\affiliation{$^4$ The  Institute  for  Research  in  Electronics  and  Applied  Physics,University  of  Maryland,  College  Park,  20742  MD,  USA}

\date{\today}

\begin{abstract} 
We study light-matter interactions in two dimensional photonic systems in the presence of a spatially homogeneous synthetic magnetic field for light. Specifically, we consider one or more two-level emitters located in the bulk region of the lattice, where for increasing magnetic field the photonic modes change from extended plane waves to circulating Landau levels. This change has a drastic effect on the resulting emitter-field dynamics, which becomes intrinsically non-Markovian and chiral,  leading to the formation of strongly coupled Landau-photon polaritons. The peculiar dynamical and spectral properties of these quasi-particles can be probed with state-of-the-art photonic lattices in the optical and the microwave domain and may find various applications for the quantum simulation of strongly interacting topological models.
\end{abstract}

\maketitle

%
%


\maketitle

The study of electronic systems in strong magnetic fields has a long tradition in condensed matter physics and led to many important discoveries such as the quantum and the fractional quantum Hall effect or flux quantization in superconducting rings \cite{Klitzing2020, FeynmanLectures}. While for a long time such effects have been restricted to charged particles, over the past years it has been shown that \emph{synthetic} magnetic fields can also be engineered for a variety of neutral systems ranging from atoms in optical lattices \cite{Goldman2016,Cooper2019} to photonic and phononic resonator arrays \cite{Ozawa2019,Carusotto2020, Peano2015}. These systems not only offer the possibility to simulate magnetic fields of unprecedented strength, but also allow us to explore novel phenomena and applications, which are not accessible with electrons. In particular, the ability to interface photons and phonons with atoms or solid-state emitters gives rise to many intriguing questions about the nature of light-matter interactions in magnetic and other topologically non-trivial environments \cite{Yao2013, Barik2018, Kim2020, Ningyuan2015, Anderson2016, Owens2018, Bello2019, Elcano2019, Longhi2019, Giorgi2020, Lemonde2019, Leonforte2020, Besedin2020, Poshakinskiy2020,Wang2020}.

In this Letter we study light-matter interactions in a 2D photonic lattice with an engineered synthetic magnetic field. Several previous works have already addressed the coupling of two-level systems to the chiral edge modes~\cite{Yao2013,Barik2018, Longhi2019,Lemonde2019,Giorgi2020}, which can be used to transport classical or quantum information in a robust and unidirectional way~\cite{Lemonde2019,Yao2013,Longhi2019,Mittal2014,Lodahl2017}.
Here we are interested in emitters coupled to the \emph{bulk} region of the photonic lattice, where the presence of magnetic fields has dramatic consequences on the dynamics of the light emission process. Intuitively, this can be understood from the fact that an emitted photon  cannot propagate away, but it is constrained to orbit around the emitter due to the effective Lorentz force~\cite{Page1930, Girvin1999}. More formally, the formation of photonic Landau levels results in a highly spiked density of states, such that even in an infinite and broad-band lattice, emitter-field interactions become intrinsically non-Markovian at \emph{all} frequencies and coupling strengths. We show that such peculiar conditions lead to a novel kind of excitations that we name Landau-photon polaritons (LPPs). By being composed of circulating \cite{footnote1} and dispersionless, but still spatially extended photons, the spectral and dynamical features of these quasi-particles can be continuously tuned from a single-mode, cavity QED type behavior to that of a many-body system of strongly interacting particles in the presence of a magnetic field. For intermediate parameter settings the hybridization of chiral photons and highly non-linear emitters results in a whole zoo of interacting magnetic lattice models, which are unprecedented in other light-matter or condensed-matter systems. This makes such systems particularly interesting for quantum simulation applications.

\begin{figure}
\centering
	\includegraphics[width=\columnwidth]{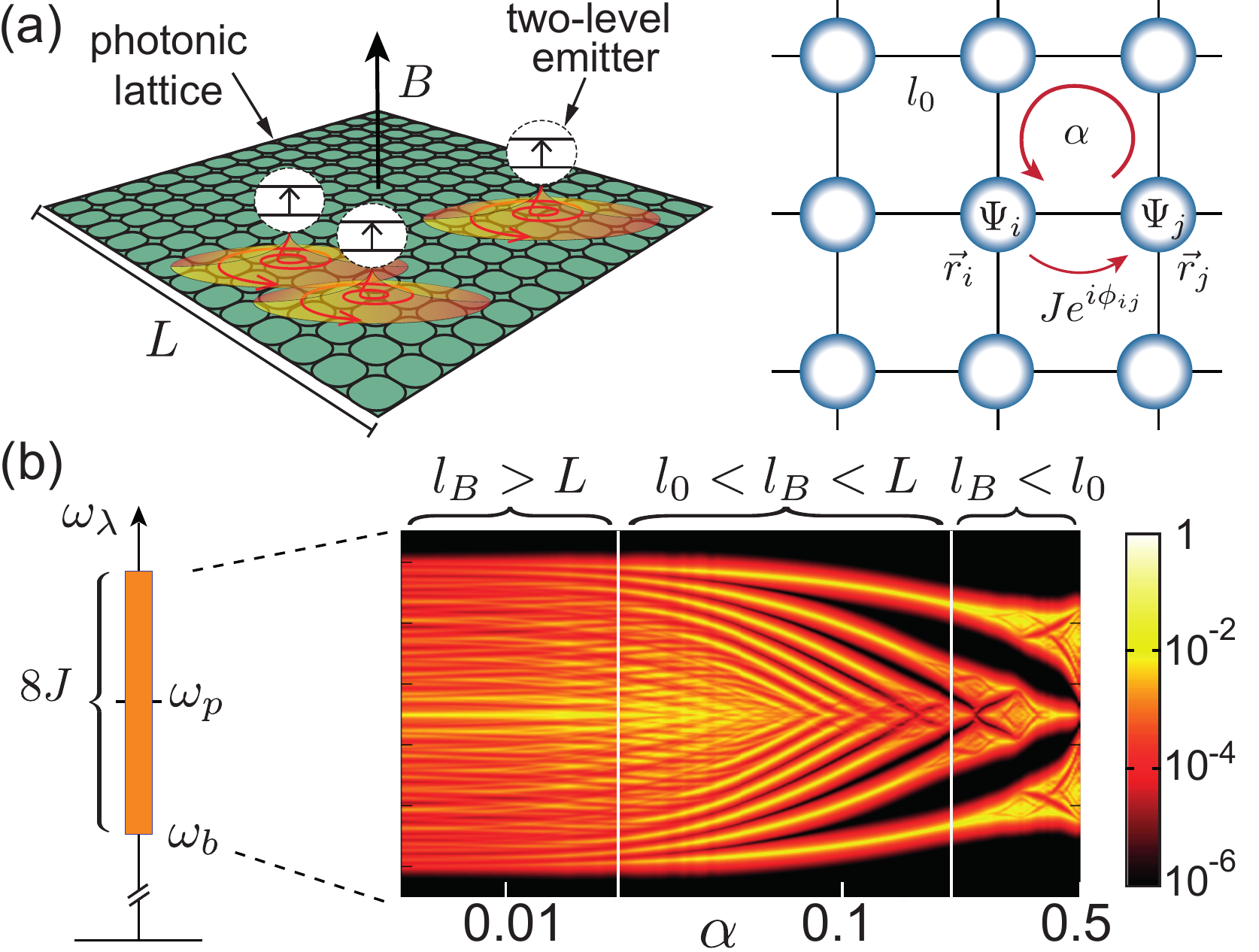}
	\caption{(a) Sketch of a system of two-level emitters coupled to a photonic lattice with a synthetic magnetic field $B$. The magnetic field is implemented by adjusting the hopping phases $\phi_{ij}$ between neighbouring lattice sites such that around each plaquette $\sum_\square \phi_{ij} =2\pi \alpha$. (b) The projected density of states, $\rho(\vec r_e,\omega)$, is plotted on a logarithmic scale (arbitrary units) as a function of $\alpha$ and for a lattice of  $M=20\times 20$ sites. For this plot, $\vec r_e/l_0=(10,10)$ and each resonance is represented by a broadened $\delta$-function with a finite width of $\gamma/J \approx 10^{-3}$. }
	\label{fig:fig1}
\end{figure}

\emph{Model.}---We consider a setup as shown in Fig.~\ref{fig:fig1}(a), where $N$ (artificial) two-level emitters with frequency $\omega_e$ are coupled to a 2D photonic resonator array of length $L$, lattice positions $\vec r_i=(x_i,y_i)$ and spacing $l_0$. Each lattice site is represented by a localized photonic mode with frequency $\omega_p$ and annihilation operator $\Psi_i\equiv \Psi(\vec r_i)$. Neighboring lattice sites are coupled via the complex tunneling amplitudes $J_{ij}=J e^{i\phi_{ij}}$. The photonic lattice is modelled by the tight-binding Hamiltonian $(\hbar =1)$
\begin{equation}
H_{\rm ph}  = \omega_p \sum_{i=1}^M  \Psi^{\dag}_i \Psi_i - J\sum_{\braket{i,j}}  \left( e^{ i\phi_{ij}} \Psi^{\dag}_i \Psi_j + {\rm H.c.} \right), 
\end{equation}
where $M=(L/l_0)^2$ is the total number of lattice sites. The Hamiltonian for the combined system is 
 \begin{equation}\label{eq:ham_main}
	\begin{split}
		H & = H_{\rm ph} +\sum_{n=1}^N   \frac{\omega_e}{2} \sigma_z^n + g \left[ \Psi(\vec r_e^{\,n}) \sigma_+^n + \Psi^\dag (\vec r_e^{\,n}) \sigma_-^n \right],
	\end{split}
\end{equation}
where the $\sigma_k^n$ are the Pauli operators for an emitter at site $\vec r_e^{\,n}$ and $g$ is the emitter-field coupling strength.

\emph{A magnetic photonic lattice.}---We are interested in the regime $N\ll M$, where a few emitters are coupled to the bulk region of a much larger photonic lattice. In a standard lattice, where $\phi_{ij}=0$,  $H_{\rm ph}$ can be diagonalised by introducing the annihilation operators $\Psi_\lambda= \sum_i f^*_\lambda(i) \Psi_i$, where the mode functions $f_{\lambda}(i)\sim e^{i\vec k_\lambda \cdot \vec r_i}$ are plane waves and the corresponding mode frequencies $\omega_{\lambda}$ form a continuous band of width $8J$ centred around $\omega_p$ [see Fig.~\ref{fig:fig1}(b)].  For $\omega_e$ within this band and $g \ll J$, an excited emitter coupled to this continuum of modes simply undergoes an exponential decay. 

Here we consider a lattice where $\phi_{ij} = \frac{e}{\hbar} \int_{\vec{r}_j}^{\vec{r}_i} \vec{A}(\vec{r})\cdot d\vec{r}$, with $\vec{A}(\vec{r}) = B(-y/2, x/2,0)$. This arrangement of phases mimics the lattice Hamiltonian for particles with charge $e$ in a homogeneous magnetic field $B$ and thus represents an equivalent synthetic magnetic field for the photons.  Such a scenario can be realized, for example, by imposing the tunneling phases through external driving fields~\cite{Peano2015,Fang2012,Roushan2017}, by engineering multi-mode lattices with effective spin-orbit interactions~\cite{Hafezi2013,Ningyuan2015,Lodahl2017,Ren2020} or by weakly hybridizing photons with magnetic materials~\cite{Anderson2016,Koch2010} to break time-reversal symmetry. See Ref.~\cite{Ozawa2019} for a more detailed discussion of those experimental techniques.

In Fig.~\ref{fig:fig1}(b) we plot the projected density of states, $\rho(\vec r_e,\omega)= \sum_\lambda |f_\lambda(\vec r_e)|^2\delta(\omega-\omega_\lambda)$, as a function of $\alpha=e\Phi/(2\pi\hbar)$, where $\Phi= B l_0^2$ is the flux enclosed in a single plaquette. This quantity captures the relevant photonic modes to which an emitter located at $\vec r_e$ is coupled to. We identify three  different regimes.  For very small $\alpha$ the magnetic length $l_{B} \simeq l_0 /(\sqrt{2\pi \alpha})$ exceeds the size of the lattice, $L$. Magnetic effects are not yet important and  $\rho(\vec r_e,\omega)$ recovers the relatively flat shape of a trivial lattice. In the opposite strong-field regime, $l_{B} \lesssim l_0$, the magnetic length is comparable to the lattice spacing and $\rho(\vec r_e,\omega)$ reproduces the fractal structure of the Hofstadter butterfly~\cite{Hofstadter1976}. 

Most relevant for the current discussion is the intermediate regime, where $l_0  < l_B < L$. In this parameter range the discreteness of the lattice is not important and we can use an effective continuum theory, where the eigenmodes $f_{\lambda}(i)\equiv\Phi_{\ell k}(\vec r_i)$ are the usual Landau orbitals~\cite{Page1930,supp},
\begin{equation}\label{eq:LandauOrbitals}
\begin{split}
& \Phi_{\ell k}(\vec r_i)\simeq \frac{l_0}{\sqrt{2\pi} l_{B}}\sqrt{\frac{\ell!}{k!} } \xi_i^{k-\ell} e^{-\frac{|\xi_i|^2}{2}} L_\ell^{k-\ell}\left(|\xi_i|^2 \right)
\end{split}
\end{equation}
with  $\xi_i=(x_i+iy_i)/\sqrt{2l^2_B}$ and $L_\ell^{k-\ell}(x)$ are generalized Laguerre polynomials. The index $\ell=0,1,2,\dots $ labels the discrete Landau levels with frequencies $\omega_{\ell}\approx \omega_b+ \omega_c (\ell+1/2)$~\cite{supp}, where  $\omega_b=\omega_p-4J$ is the frequency of the lower band edge and  $\omega_c= 4\pi \alpha J$ is the cyclotron frequency. The second index $k=0,1,2 \ldots$ labels the $\sim \alpha M$ degenerate modes within each band. 
Clearly, both the transformation from a continuous to a discrete spectrum and the localization of the photonic eigenmodes will strongly affect the physics of light-matter interactions in such a synthetic magnetic environment.

\emph{Single-emitter dynamics.}---We first consider the case of a single emitter located at position $\vec r_e$ in the bulk of the lattice. The emitter is initially prepared in its excited state and the system's wavefunction can be written as $|\psi\rangle(t)= e^{-i\omega_{e} t}[c_e(t) \sigma_+ + \sum_i \varphi(\vec r_i,t) \Psi^\dag(\vec r_i)]|g\rangle|{\rm vac}\rangle$, where $c_e(t)$ is the emitter amplitude and $\varphi(\vec r_i,t)$ the photon wavefunction. From this ansatz we obtain
\begin{equation}\label{eq:eq_retarded_single_ex_many_atoms}
\dot{c}_e(t) = - g^2  \int_0^t ds \,G(t-s,\vec r_e, \vec r_e) c_e(s) e^{i\omega_e(t-s)} ,
\end{equation} 
where $G(\tau ,\vec r_i, \vec r_j)= \langle {\rm vac}| \Psi(\vec r_i,\tau )\Psi^\dag(\vec r_j,0)|{\rm vac}\rangle=\sum_\lambda f_\lambda(\vec r_i) f^*_\lambda(\vec r_j) e^{-i\omega_\lambda \tau}$ is the photonic Green's function.

\begin{figure}
\centering
	\includegraphics[width=\columnwidth]{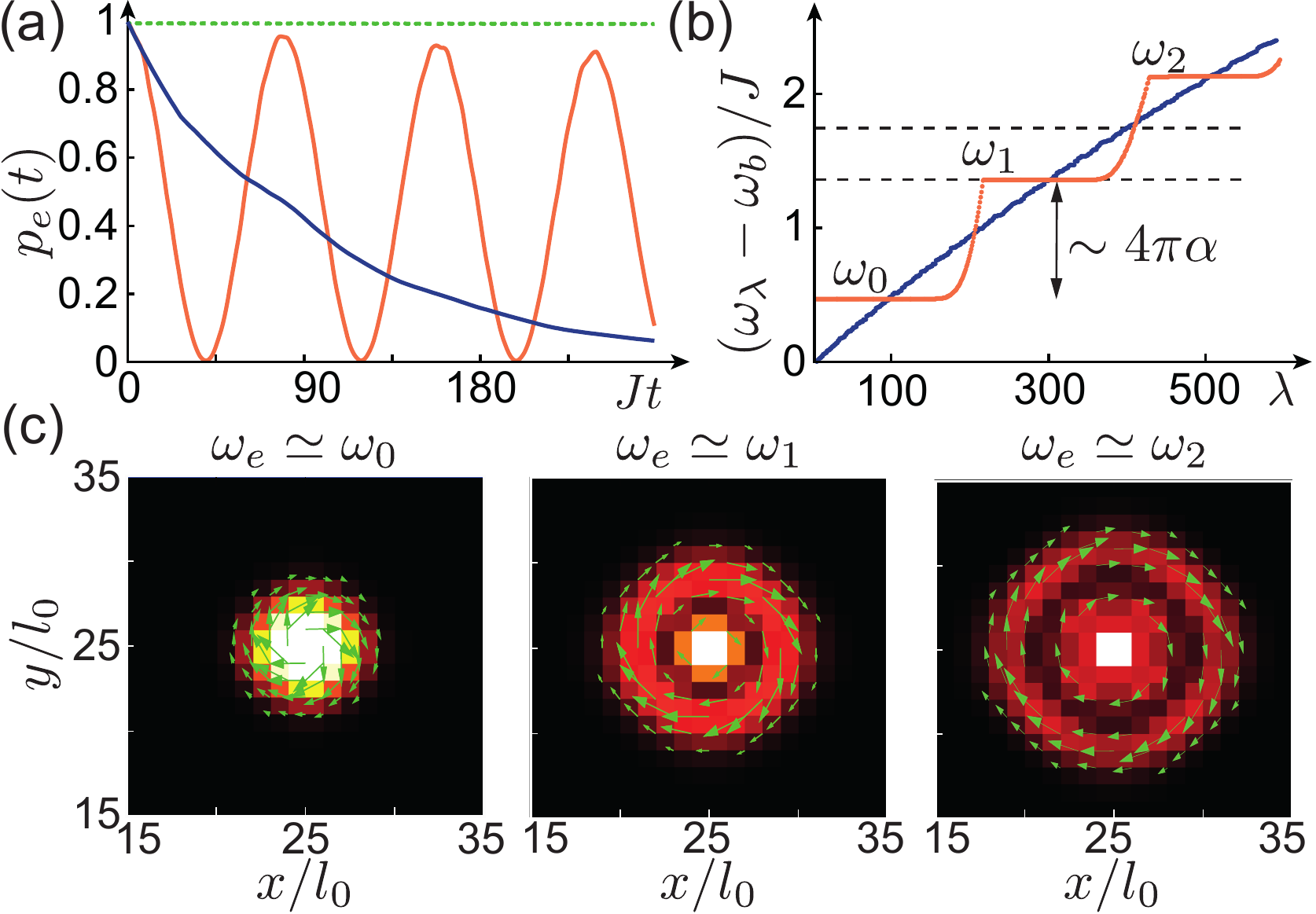}
\caption{(a) Evolution of the excited-state population, $p_e(t)$, of an emitter located at $\vec r_e/l_0=(25,25)$ in a lattice of $50\times 50$ sites. The parameters are $\alpha=0$ and $\delta_e/J = 1.35$ (blue line), $\alpha=0.08$ and $\delta_e/J = 1.35$ (orange line), and $\alpha=0.08$ and $\delta_e/J = 1.76$ (green dashed line). (b) Plot of the lowest eigenfrequencies $\omega_\lambda$ of the two photonic lattices as used for the simulation shown in blue and orange in (a). The dashed black lines indicate the corresponding emitter's frequencies. 
(c) Photon density, $|\varphi(\vec r_i,t_\pi)|^2$, combined with the profile of the photon current, $\langle \vec{j}_p\rangle(\vec r_i,t_\pi)$, at time $t_\pi=\pi/(2\Omega)$, for $\alpha=0.08$ and $
\omega_e=\omega_{\ell=0,1,2}$. For all plots $g/J=0.14$ and for each lattice site in the bulk (on the edge) a photon decay rate of $\gamma_p/J = 4\times 10^{-4}$ ($\gamma_{\rm edge}/J \sim 10^{-1}$) has been introduced~\cite{supp}. }
	\label{Fig2:SingleAtom}
\end{figure}

 In Fig.~\ref{Fig2:SingleAtom}(a) we show the evolution of the excited-state population, $p_e(t)=|c_e(t)|^2$, for different $\alpha$ and different detunings from the band edge,  $\delta_e= \omega_e-\omega_b$. For $\alpha=0$ and $M\rightarrow \infty$ the Green's function $G(\tau, \vec r_e,\vec r_e)$ is represented by a mode continuum and decays on a short time scale, $J^{-1}$. It is then valid to make a Markov approximation and, consistent with a numerical simulation of the full wavefunction $|\psi\rangle(t)$, we obtain an exponential decay of $p_e(t)$ with a rate  $\Gamma\simeq 2\pi g^2 \rho(\vec r_e,\omega_e)\approx g^2/(2J)$ \cite{supp}. For $\alpha\neq 0$ the situation is very different and depending on $\omega_e$ we observe either no decay at all or coherent oscillations. This behaviour can be understood from the exact spectrum of $H_{\rm ph}$ plotted in Fig.~\ref{Fig2:SingleAtom}(b). It exhibits discrete plateaus at frequencies $\omega_\ell$ connected by a sparse set of intermediate modes representing the edge states. Since an emitter in the bulk does not see the edges, whenever $|\omega_e-\omega_\ell|\gtrsim g$ there are no available modes to couple to and the emitter remains frozen in the excited state.
 
The situation is very different when $\omega_e\simeq \omega_\ell$, in which case the emitter couples to a flat band without dispersion. We can then project the Green's function on the resonant Landau level and obtain $G(\tau ,\vec r_i, \vec r_j)\simeq G_{\ell}(\vec r_i, \vec r_j)e^{-i\omega_\ell \tau}$, where 
\begin{equation}
G_{\ell}(\vec r_i, \vec r_j)\simeq   \sqrt{\alpha} e^{ i\theta_{ij} }   \Phi_{\ell \ell} (\vec r_i-\vec r_j)
\end{equation}  
and $\theta_{ij} =-(x_i y_j-x_j y_i)/(2 l_{B}^2)$ \cite{supp}.  
Under this approximation,  Eq.~\eqref{eq:eq_retarded_single_ex_many_atoms} can be converted into a second-order differential equation, $\ddot c_e= - \Omega^2 c_e$. Here 
\begin{equation}
\Omega= \sqrt{\alpha} g
\end{equation} 
is the vacuum Rabi frequency, which has the same value for all Landau levels.  The predicted Rabi oscillations, $p_e(t)=\cos^2(\Omega t)$, are exactly reproduced by the full numerical simulation, keeping in mind that in Fig.~\ref{Fig2:SingleAtom}(a) we have included a finite loss rate $\gamma_p$ for all photons to describe a realistic scenario. For the photon wavefunction we obtain
\begin{equation}
\varphi(\vec r_i,t) = -i\frac{\sin(\Omega t)}{\sqrt{\alpha}} G_{\ell}(\vec r_i, \vec r_e),
\end{equation}
i.e., at time $t_\pi=\pi/(2\Omega)$ the excitation is fully converted into a circulating photon in the Landau orbital $\sim \Phi_{\ell\ell} (\vec r_i-\vec r_e)$, centered around the emitter. This is shown in Fig.~\ref{Fig2:SingleAtom}(c) in terms of the density, $|\varphi(\vec r_i,t_\pi)|^2$, and the photon-current profile, $\langle \vec{j}_p\rangle(\vec r_i,t_\pi)$ \cite{supp}. Note  that all  these results are independent of the gauge for $\vec A$ and the chosen Landau basis in Eq.~\eqref{eq:LandauOrbitals}, which depends explicitly on the origin of the coordinate system. However, $G_{\ell}(\vec r_i, \vec r_j)$ still includes a gauge-dependent phase factor, $\theta_{ij}$, which will become important in the following.

\emph{Landau-photon polaritons.}---Let us now extend these results to multiple emitters, still focusing on the regime $\omega_c\gg g$, where the emitters couple dominantly to a single Landau level. In this case each emitter only interacts with photons in the  orbital centered around its location, $\Phi_{\ell\ell} (\vec r_i-\vec r^{\,n}_e)$. The photons themselves do not evolve, because there is no dispersion. These special conditions allow us to restrict the dynamics of the whole lattice to a reduced set of modes with bosonic operators
\begin{equation}
B_{\ell n} = \sum_{m=1}^N (K^{-1})_{n m}  \sum_i  G_{\ell}(\vec r_e^{\,m}, \vec r_i)\Psi(\vec{r}_i).
\end{equation}  
Here, the $N\times N$ matrix $K$ satisfies $(KK^\dag)_{n m} = G_\ell(\vec r_e^{\,n},\vec r_e^{\,m})$~\cite{supp}, which ensures that the $B_{\ell n}$ form an orthogonal set of modes with $[ B_{\ell n}, B_{\ell m}^{\dag}]=\delta_{nm}$. Projected onto these modified Landau orbitals, we obtain the effective Hamiltonian
\begin{equation}\label{eq:HLPP}
\begin{split}
H^{(\ell)}_{\rm LPP} =\,& \omega_{\ell} \sum_{n=1}^N  B_{\ell n}^\dag B_{\ell n}  +  \frac{\omega_e}{2}\sum_{n=1}^N    \sigma_z^n\\
+& g\sum_{n,m=1}^N \left(  \sigma_+^n K_{nm} B_{\ell m}  + B_{\ell m}^\dag K_{nm}^* \sigma_-^n\right).
\end{split} 
\end{equation}
It describes the full nonlinear dynamics of LPPs, which are the quasi-particles formed by the coupling of two-level emitters to photons in a single Landau level. This  model generalizes the dressed emitter-emitter interactions introduced in~\cite{peppino2016} and holds even in the presence of a finite bandwidth $J_\ell$ or local frequency disorder $\Delta \omega_p$~\cite{Halperin1982}, as long as $\omega_c \gg g\gg J_\ell,\Delta \omega_p$~\cite{supp}.
Importantly, $H^{(\ell)}_{\rm LPP}$ only involves $N$ independent photonic modes, i.e., considerably fewer than the number of lattice sites. This makes few-excitation physics numerically tractable, which usually is not possible in 2D waveguide QED systems. 
In Fig.~\ref{Fig3:LPP}(a) we show the single- and two-excitation spectrum of $H^{(\ell=1)}_{\rm LPP}$ for $N=3$ equally spaced emitters with $|\vec r_e^{\,n}-\vec r_e^{\,m}|=d$ and assuming resonance conditions, $\omega_e=\omega_{\ell=1}$.

\begin{figure}
\centering
	\includegraphics[width=\columnwidth]{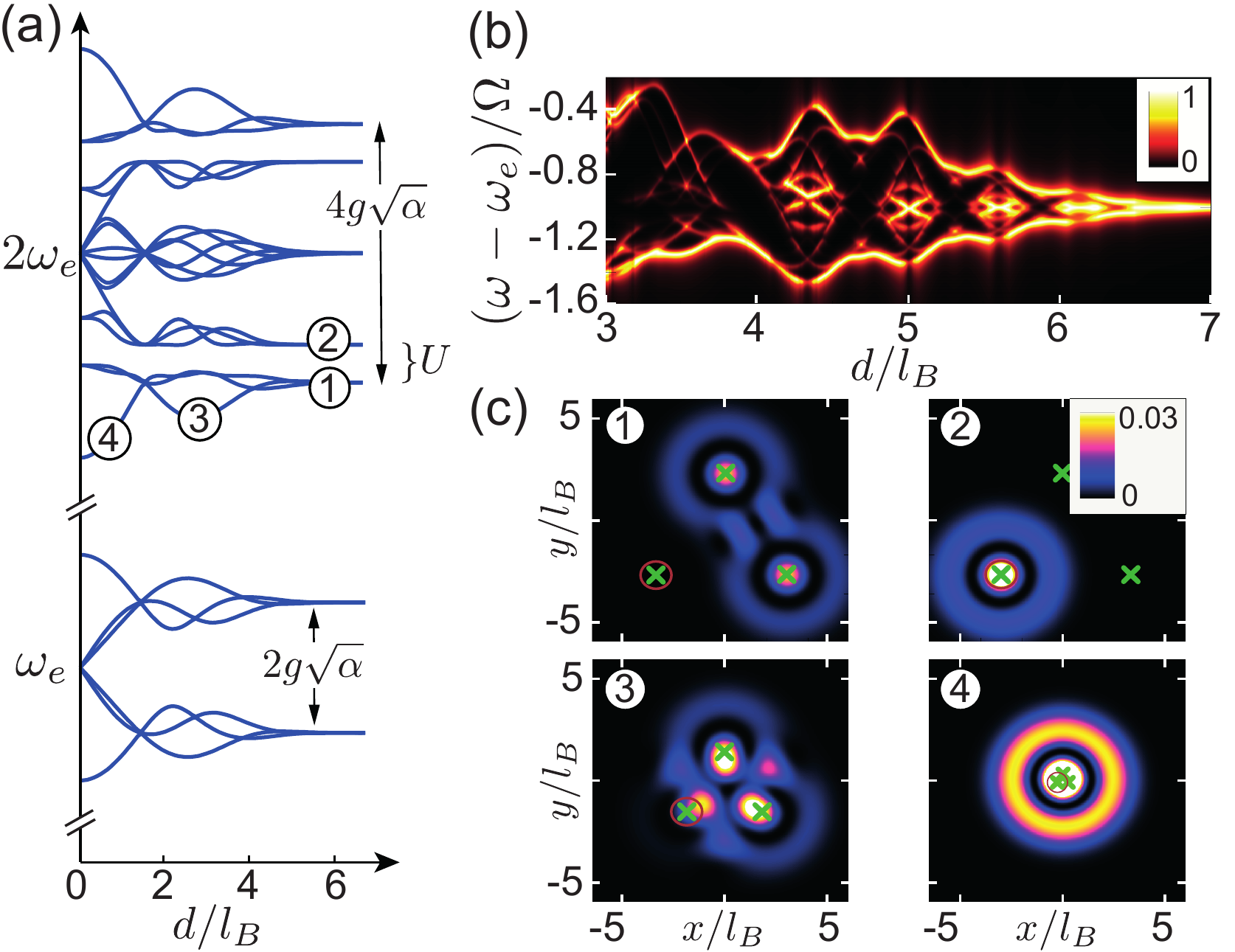}
	\caption{(a) The spectrum of $H^{(\ell)}_{\rm LPP}$ in the single- and two- excitation sector for $N=3$ equidistant emitters with varying spacing $d$ and for $\ell=1$ and $\omega_e = \omega_1$. (b) Zoom of the lower polaritonic band of the emitter's excitation spectrum, $\mathcal{S}^{n}_e(\omega)$, for a $N=4\times 4$ square lattice of emitters with open boundaries, where $\vec r_e^{\, n}$ is the location one of the four inner emitters (the full spectrum is reflection symmetric around $\omega_e$). For this plot, $\alpha=0.08$, $\gamma_e/\Omega=0.02$ and $\ell = 3$. The color scale is normalised to the maximum value. (c) Plot of the two-photon correlation function $C(\vec r_i,\vec r_e^{\,1})$ for the different two-photon eigenstates indicated in (a). The green crosses represent the emitters position, and the red circle marks the reference emitter's position $\vec r_e^{\,1}$ .}
	\label{Fig3:LPP}
\end{figure}

For a single excitation we obtain an upper and a lower polariton branch, which split into subbands of frequencies
\begin{equation}
\omega_{\ell,\nu}^\pm  = \omega_e \pm  \Omega \sqrt{1  + e^{-\frac{d^2 }{4l_B^2}} L_\ell^0 \left( \frac{d^2}{2l_B^2}\right) \lambda_{\nu} },
\end{equation}
where $\lambda_{\nu=1,2,3}= 2 \cos[(\theta_\triangle+2\pi \nu)/3]$ and  $\theta_{\triangle} = \theta_{12} + \theta_{23} + \theta_{31}=e  B A_{\triangle}/\hbar$ is the normalized flux through the area $A_{\triangle}$ enclosed by the three emitters. 
For widely separated emitters, each emitter supports an independent pair of upper/lower polariton states with the same Rabi splitting $2\Omega$. When the spacing $d$ between emitters is reduced, the photonic wavefunctions start to overlap and each polariton manifold splits into three branches, similar to the formation of binding and anti-binding orbitals in molecules.
The dependence on both the enclosed flux as well as on the shape of the Laguerre polynomials quantifying the wavefunction
overlap make the spectra of LPPs rather complex.
The $\lambda_\nu$ reflect the characteristic eigenvalue structure of
a three-site lattice in a magnetic field and the symmetry between the upper and lower polaritons is a consequence of the resonant coupling of the emitters to a single and degenerate Landau level. For any $\theta_\Delta \neq n\pi$, the left- and right-circulating polariton modes are no longer degenerate, which indicates chirality of LPP propagation \cite{Koch2010}.

The physics is even richer in large lattices, as exemplified in Fig.~\ref{Fig3:LPP}(b) for a square lattice of $N = 4\times4$ emitters.
This plot shows the  lower part of the emitter's excitation spectrum,
\begin{equation}\label{eq:Spectrum}
\mathcal{S}^{n}_e(\omega) = 
\left| \braket{G | \sigma_-^n  \frac{1}{H - \omega  - i\frac{\gamma_e}{2}\sum_{m}\sigma_+^m\sigma_-^m }\sigma_+^n | G} \right|^2,
\end{equation}
where $|G\rangle$ is the ground state and $\gamma_e$ is the bare decay rate of the emitters. 
The repetitive features in this spectrum can be understood in terms of a Harper-Hofstadter model with a flux $\sim d^2/l_B^2$ per plaquette. This spectrum can be directly obtained by detecting the light scattered from one weakly driven emitter.

Let us move to the multiple excitation case.
It is well-known that in a single-mode system, the Jaynes-Cummings interaction gives rise to an effective repulsion,  $U=\Omega(2-\sqrt{2})$, between two polaritons. This interaction can also be clearly identified in Fig.~\ref{Fig3:LPP}(a), where at large distance $d$ the lowest three eigenstates in the two-excitation sector are separated by $U$ from the next three levels. The difference between these two sets of polaritonic states can be visualized in terms of the two-photon correlation function, $C(\vec r_i,\vec r_j)=\langle \Psi^\dag(\vec r_j) \Psi^\dag (\vec r_i) \Psi (\vec r_i) \Psi(\vec r_j)\rangle/\langle  \Psi^\dag (\vec r_j) \Psi (\vec r_j) \rangle $, plotted in Fig.~\ref{Fig3:LPP}(c). 
For $d\gg l_B$, the energetically lowest states exhibit strong anti-bunching, $C(\vec r_i,\vec r_j)\simeq 0$ for $|\vec r_i-\vec r_j|\lesssim l_{B}$, reminiscent of a Laughlin wavefunction, where particles avoid each other. For the interacting states we obtain $C(\vec r_e^{\,n},\vec r_e^{\, m})\simeq 0$ for $n\neq m$, meaning that both photons occupy the same orbital. At smaller distances, the kinetic energy, i.e., the overlap between orbital states becomes more relevant and anti-bunching gradually disappears with details depending on the enclosed magnetic flux, $\theta_{\triangle}$. For $d \leq l_B$, the emitters couple identically to the field, such that the interactions become fully collective and the spectrum converges to that of a single-mode Tavis-Cummings model \cite{Tavis1968}.

\emph{Chiral dipole-dipole interactions and effective flat-band models.}---The situation is most transparent and intriguing when the emitters are sufficiently detuned from the nearest Landau level, $|\omega_e-\omega_\ell|\gg g$. In this case they are only weakly dressed by the photons, which gives rise to effective dipole-dipole interactions of the form $H_{\rm eff} =\sum_{n,m}  \left(\tilde J_{nm} \sigma_+^n \sigma_-^m + {\rm H.c.} \right) $. Here  
\begin{equation}
\tilde J_{nm} \simeq  \frac{g^2}{\omega_e-\omega_\ell} |G_{\ell}(\vec r_e^{\,n},\vec r_e^{\,m})| e^{ i \theta_{nm}}, 
\end{equation}
are complex hopping amplitudes, which inherit the magnetic features from the photons.  Therefore, also in this almost decoupled limit, dipole-dipole interactions between $N\geq 3$ emitters depend sensitively on the magnetic flux, which can lead to a fully chiral transport of excitations. As illustrated in Fig.~\ref{Fig4:AtomAtom}(a), a single excitation flows in the clockwise direction, 
while two excitations lead to an anti-clockwise dynamics for their relative hole \cite{Roushan2017}.

\begin{figure}
\centering
	\includegraphics[width=\columnwidth]{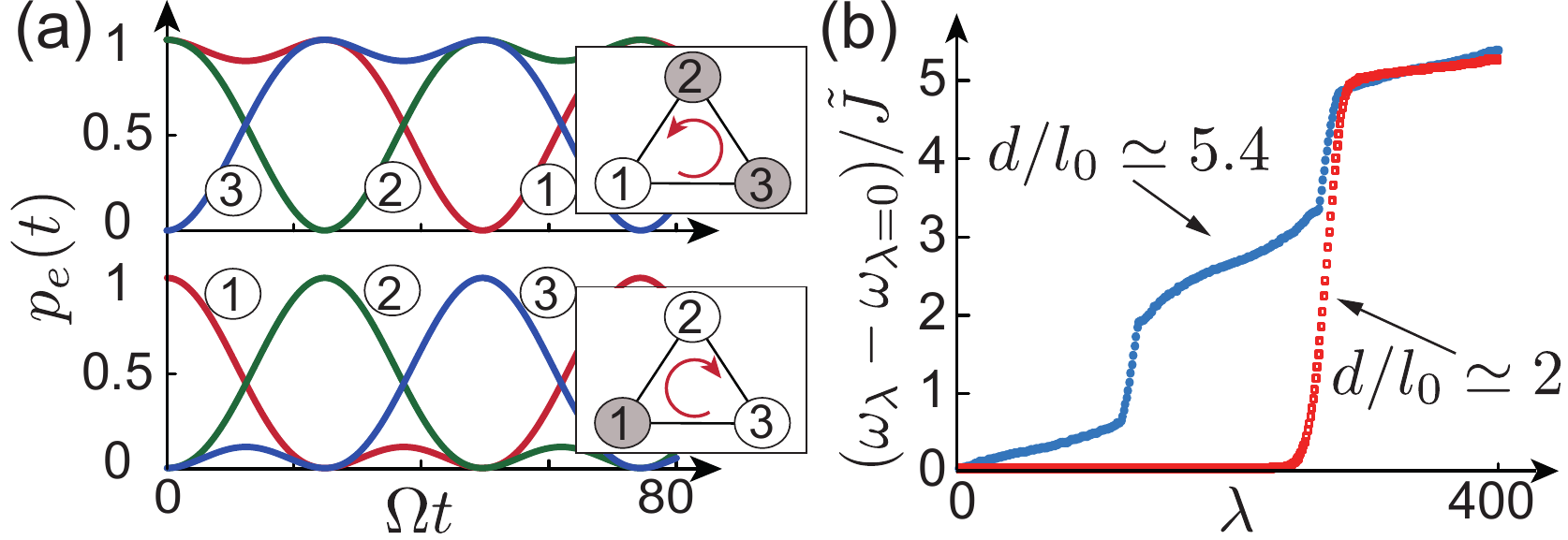}
	\caption{(a) Evolution of the excited state populations $p_e^n(t)$ of $N =3$ emitters arranged in a triangle of length $d/l_0=4$. For this plot $\ell = 0$, and $\alpha=1/(16\sqrt{3}) \approx 0.036$, such that the enclosed effective flux is $\theta_{\triangle}\simeq \pi/2$ and the dipole-dipole interactions become fully chiral (see \cite{supp} for details). In the upper panel the initial state contains two excitations in emitter 1 and 2. In the lower panel the initial state contains just one excitation in emitter 1. (b) Single-excitation spectrum of $H_{\rm eff}$ for a square lattice of $20\times 20$ emitters and normalized to the nearest-neighbor coupling strength $\tilde J=|\tilde J_{12}|$. The two spectra are obtained for the spacings $d/l_0 = 2$ ($\alpha_{\rm eff} = 0.32$) and $d/l_0 = 5.39$ ($\alpha_{\rm eff} = 2.32$) and in both cases $\alpha=0.08$ and $\ell=0$ has been assumed.}
	\label{Fig4:AtomAtom}
\end{figure}

More generally, the effective Hamiltonian $H_{\rm eff}$ can be viewed as a magnetic lattice model for hard-core bosons, 
with various additional interesting features. Analogously to Fig. \ref{Fig3:LPP}, the magnetic flux associated with the phases $\theta_{ij}$ depends on the emitter's arrangement and can be considerably enhanced, i.e., $\alpha_{\rm eff}=\alpha  (d/l_0)^2$ for a square lattice. Further,  tunneling is no longer constrained to nearest neighbors and depending on the spacing, the lattice geometry and the Landau-level index $\ell$, a whole zoo of magnetic models with different band-structures and field strengths can be realized. For example, in Fig.~\ref{Fig4:AtomAtom}(b) we show the single-excitation spectrum of $H_{\rm eff}$ for a square lattice of emitters for two different spacings, but equivalent effective field strengths.  In the first case, only nearest-neighbor couplings are relevant and we recover the regular Hofstadter butterfly with $\alpha_{\rm eff}\approx 2.32$ (which is equivalent to $\alpha_{\rm eff}\approx 0.32$). In the second example, long-range hoppings are important and the spectrum of the bulk modes becomes essentially flat. This situation is reminiscent of the spectrum of the Kapit-Muller Hamiltonian \cite{Kapit2010}, a prototype toy model for strongly interacting magnetic systems. Interestingly, such abstract models arise very naturally from the coupling of emitters to a magnetic photonic reservoir.

We emphasize that the strong coupling of superconducting qubits to arrays of microwave resonators in the regime $g\gg \gamma_e, \gamma_p$~\cite{Liu2016,Kim2020,Besedin2020} as well as the implementation of synthetic fields in small~\cite{Roushan2017} and large two-dimensional~\cite{Owens2018} photonic lattices have already been demonstrated. A combination of these techniques is sufficient to probe all the characteristic properties of LPPs with state-of-the-art parameters~\cite{supp}. With further developments, similar experiments should also become possible with atoms or solid-state emitters coupled to topological lattices in the optical regime~\cite{Hafezi2013, Mittal2014, Mittal2018, Barik2020}.

\emph{Conclusions.}---In summary, we have shown how the presence of synthetic magnetic fields changes the physics of light-matter interactions in the bulk of 2D photonic lattices. For moderate magnetic fields this physics can be very accurately described in terms of LPPs, which share the nonlinearity of the matter component and the chiral properties of Landau photons. In the many emitter case, our platform naturally allows the quantum simulation of various interaction-dominated topological systems, which do not appear in electronic systems with only nearest-neighbor interactions.

\acknowledgements
We thank Giuseppe Calajo and Francesco Ciccarello for stimulating discussions. This work was supported by the Austrian Academy of Sciences (\"OAW) through a DOC Fellowship (D.D.) and by the Austrian Science Fund (FWF) through the DK CoQuS (Grant No.~W 1210) and Grant No. P31701 (ULMAC). We acknowledge financial support from the European Union FET-Open grant ``MIR-BOSE” (n.737017), from the H2020-FETFLAG-2018-2020 project ``PhoQuS” (n.820392), from the Provincia Autonoma di
Trento, from the Q@TN initiative, from Google via the quantum NISQ award,
and from AFOSR MURI FA9550-19-1-0399 and ARO MURI W911NF-15-1-0397.

\onecolumngrid
\clearpage
\setcounter{equation}{0}
\setcounter{figure}{0}
\setcounter{table}{0}
\setcounter{page}{6}
\makeatletter
\renewcommand{\theequation}{S\arabic{equation}}
\renewcommand{\thefigure}{S\arabic{figure}}
\renewcommand{\bibnumfmt}[1]{[S#1]}
\renewcommand{\citenumfont}[1]{S#1}
\begin{center}
        \textbf{\large Supplementary material for: \\ Light-matter interactions in synthetic magnetic fields: Landau-photon polaritons}
\end{center}

\title{Supplementary material for: \\ Light-matter interactions in synthetic magnetic fields: Landau-photon polaritons}

\author{Daniele De Bernardis$^1$, Ze-Pei Cian$^2$, Iacopo Carusotto$^3$, Mohammad Hafezi$^{2,4}$, Peter Rabl$^1$}
\affiliation{$^1$Vienna Center for Quantum Science and Technology, Atominstitut, TU Wien, 1040 Vienna, Austria}
\affiliation{$^2$ Joint  Quantum  Institute,  College  Park,  20742  MD,  USA}
\affiliation{$^3$ BEC-CNR-INFM and Dipartimento di Fisica, Universit`a di Trento, I-38050 Povo, Italy}
\affiliation{$^4$ The  Institute  for  Research  in  Electronics  and  Applied  Physics,University  of  Maryland,  College  Park,  20742  MD,  USA}

\date{\today}

\maketitle

\section{Continuum limit}
\label{app:cont_model}
The magnetic Hamiltonian $H_{\rm ph}$ for the photonic lattice is quadratic in the field operators and can be written in a diagonal form as
\begin{equation}
H_{\rm ph}= \sum_\lambda \omega_\lambda \Psi_\lambda^\dag \Psi_\lambda, \qquad {\rm where}\qquad [\Psi_\lambda,\Psi_{\lambda'}^\dag]=\delta_{\lambda,\lambda'}.
\end{equation}
By making the ansatz $|\varphi_\lambda\rangle= \Psi^\dag_\lambda|{\rm vac}\rangle = \sum_i f_\lambda(\vec r_i) \Psi^\dag(\vec r_i)|{\rm vac}\rangle$ for a single-photon eigenstate of $H_{\rm ph}$, the eigenfrequencies $\omega_\lambda$ and the corresponding mode functions $f_\lambda(\vec r)$ can be derived from the eigenvalue equation
\begin{equation}\label{eq:EVequation}
(\omega_\lambda-\omega_p)  f_\lambda(\vec r_i ) = - J \left[e^{ -i\phi_x} f_\lambda ( \vec r_i+ \vec e_x)  +e^{ i\phi_{x}} f_\lambda ( \vec r_i- \vec e_x) +e^{ -i\phi_{y}} f_\lambda( \vec r_i+ \vec e_y)+ e^{ i\phi_{y}}f_\lambda ( \vec r_i- \vec e_y)  \right].
\end{equation}
Here $\vec e_{x,y}$ are the two lattice unit vectors and we introduced the short notation 
\begin{equation}
\phi_{x,y} = \frac{e}{\hbar} \int_{\vec{r}_i}^{\vec{r}_i +  \vec e_{x,y}} \vec{A}(\vec{r})\cdot d\vec{r} \simeq \frac{e}{\hbar}\vec{A}(\vec{r}_i)\cdot \vec e_{x,y}.
\end{equation}
In the last step we have assumed that the vector potential doesn't vary considerably over the extent of one lattice site. If we restrict ourselves to moderate fields and low-frequency excitations we can also replace $f_\lambda(\vec r)$ by a continuous function and perform a Taylor expansion, 
\begin{equation}
f_\lambda ( \vec r_i+ \vec e_x)\simeq f_\lambda( \vec r_i)+ l_0 \frac{\partial }{\partial x} f_\lambda ( \vec r_i)+\frac{l_0^2}{2} \frac{\partial^2 }{\partial x^2} f_\lambda( \vec r_i).
\end{equation}  
Then, up to second order in $l_0$, the terms on the right hand side of Eq.~\eqref{eq:EVequation} can be approximated by
\begin{equation}
- J \left[e^{ - i\phi_x} f( \vec r_i+ \vec e_x)  +e^{ i\phi_{x}} f_\lambda ( \vec r_i- \vec e_x)\right] \simeq -2J f_\lambda (\vec r_i) -J l_0^2\left[ \frac{\partial }{\partial x} - i \frac{e}{\hbar} A_{x}(\vec{r}_i) \right]^2 f_\lambda (\vec r_i) +O(l_0^3).
\end{equation}
Therefore, we end up with a partial differential equation 
\begin{equation}\label{eq:EVcontinuum}
\hbar (\omega_\lambda-\omega_b)  f(\vec r ) = \frac{1}{2m} \left[-i\hbar \vec \nabla - e \vec A(\vec r)\right]^2   f(\vec r ),
\end{equation}
where $\omega_b=\omega_p-4J$ and $m=\hbar/(2J l_0^2)$ is the effective mass in the lattice.

\subsection{Landau orbitals}
Equation~\eqref{eq:EVcontinuum} is the Schr\"odinger equation for a charged $e$ particle in a magnetic field, for which the eigenfunctions are the well-known Landau orbitals, $f_\lambda(\vec r)\equiv \tilde \Phi_{\ell k}(\vec r)$. In this work we use the symmetric gauge, $\vec{A} = B(- y/2, x/2,0)$, where \cite{Page1930S}
\begin{equation}\label{eq:LOcontinuum}
 \tilde \Phi_{\ell k}(\vec r)= \frac{1}{\sqrt{2\pi l_{B}^2}}\sqrt{\frac{\ell!}{k!} } \xi^{k-\ell} e^{-\frac{|\xi|^2}{2}} L_\ell^{k-\ell}\left(|\xi|^2 \right). 
\end{equation}
Here $L_\ell^{k-\ell}(x)$ are generalised Laguerre polynomials, $l_{B}= \sqrt{\hbar/eB} $ and $\xi=(x+iy)/\sqrt{2l^2_B}$. The wavefunctions depend on two indices, $\ell$ and $k$. The index $\ell=0,1,2,\dots$ labels the Landau levels with frequencies $\omega_\ell =\omega_b +\omega_c (\ell+1/2)$, where 
\begin{equation}
\omega_c = \frac{eB}{m} = 4\pi \alpha J.
\end{equation} 
Each of these Landau levels contains a large number of degenerate sublevels, which are labeled by the second quantum number $k=0,1,2, ..., k_{\rm max}$ \cite{Girvin1999S}. For a finite system the level of degeneracy can be estimated by $k_{\rm max}\approx \alpha M  \gg1$ (where $M$ is the total number of lattice sites). For all our analytic calculations we take the limit $k_{\rm max}\rightarrow \infty$, which is a good approximation for moderate field strengths and sufficiently far away from the boundaries. 

Note that the Landau orbitals given in Eq.~\eqref{eq:LOcontinuum} denote the wavefunctions in the continuum. They are normalized to 
\begin{equation}
\int d^2 r   \, \tilde \Phi^*_{\ell k}(\vec r) \tilde \Phi_{\ell' k'}(\vec r) =\delta_{\ell\ell'} \delta_{kk'}.
\end{equation}
The corresponding normalized wavefunctions on the lattice, as given in Eq. (3)
in the main text, can then be obtained by identifying $ \Phi_{\ell k}(\vec r_i)=   \tilde \Phi_{\ell k}(\vec r=\vec r_i) l_0$. These wavefunctions have the important property that
\begin{equation}
\Phi_{\ell \ell}(\vec r=0) = \sqrt{\alpha}.
\end{equation}
This implies that the coupling between a single emitter and a single photon is independent of $\ell$.

\subsection{Lattice corrections to the Landau levels energy}
The continuum approximation is strictly valid only in the limit $\omega_c/J\sim \alpha \rightarrow 0$. While for the parameter regimes considered in this work this approximation still gives very accurate predictions for the wavefunctions, there are notable corrections to the frequencies $\omega_\ell$. To derive the lowest-order corrections to the equally spaced Landau levels, it is more convenient to use the so-called Harper equation \cite{Hofstadter1976S}, which is just the discrete single particle Schr\"odinger equation from above, but expressed in the Landau gauge, where $\vec A=B(0,x,0)$. This equation reads
\begin{equation}\label{eq:Harper}
-J[f_\lambda ( \vec r_j+ \vec e_x) + f_\lambda ( \vec r_j- \vec e_x)] - 2J\cos \left( 2\pi \alpha j - k_y \right)f_\lambda ( \vec r_j) = (\omega_{\ell}-\omega_p) f_\lambda ( \vec r_j),
\end{equation}
where $k_y$ labels the momentum in the $y$-direction, which is a good quantum number in the Landau gauge and $f_\lambda ( \vec r_j)= \chi_\lambda ( x_j)e^{i k_y y_j}$. Different values of $k_y$ only lead to a translation of the wavefunction and for a sufficiently large lattices  we can take $k_y=0$ without loss of generality. Then, following Ref.~\cite{Harper2014S}, we replace $\chi_\lambda ( x_j)$ by a continuous, slowly varying function and expand both the cosine and the discrete derivative in Eq.~\eqref{eq:Harper} up to fourth order in $l_0$, i.e., 
\begin{equation} 
-J[\chi_\lambda ( x+ l_0) + \chi_\lambda ( x- l_0) ] \simeq -2J \chi_\lambda( x) - J l_0^2 \frac{\partial^2}{\partial x^2} \chi_\lambda (x) - \frac{J l_0^4}{12} \frac{\partial^4}{\partial x^4}\chi_\lambda (x)
\end{equation}
and, using $x = jl_0$ and $2\pi\alpha=(l_0/l_B)^2$,
\begin{equation} 
-2J\cos \left( 2\pi \alpha j \right)\chi_\lambda (x)\simeq \left[ -2J  + J \frac{l_0^2}{l_B^4} x^2 - J \frac{l_0^4}{12 l_B^8} x^4 \right]\chi_\lambda (x).
\end{equation}
With the definitions introduced above we then obtain the Schr\"odinger equation 
\begin{equation}
\hbar (\omega_{\ell}-\omega_p-4J) \chi_\lambda (x)= \left[- \frac{\hbar^2}{2m} \frac{\partial^2}{\partial x^2}+ \frac{1}{2} m \omega_c^2 x^2 \right] \chi_\lambda (x) - \frac{1}{48\hbar J}  \left[ \frac{\hbar^4}{m^2} \frac{\partial^4}{\partial x^4}+  m^2 \omega_c^4 x^4 \right] \chi_\lambda (x).
\end{equation}
The first term on the right hand side is just the Hamiltonian of a harmonic oscillator, from which we recover the the equidistant Landau levels, $\omega_\ell=\omega_b +\omega_c(\ell+1/2)$.  The second term contains the lowest order corrections to the purely harmonic oscillator, which are fourth order in the momentum and the position operators. By including these corrections in perturbation theory we obtain the more accurate Landau spectrum~\cite{Harper2014S}
\begin{equation}
\omega_{\ell} \simeq \omega_b + \omega_c \left( \ell + \frac{1}{2}\right)- \frac{\omega_c^2 }{32J}(2\ell^2+2\ell+1).
\end{equation}
%
For example, based on this formula, the gap between the two lowest Landau levels is given by
\begin{equation}
\omega_1 -\omega_0 \approx 4\pi \alpha J \left( 1 - \frac{\pi}{2} \alpha \right). 
\end{equation}
If we assume a value of $\alpha=0.08$, as in many examples in the main text, we find
\begin{equation}
\frac{\omega_1 -\omega_0}{J} \approx 0.874.
\end{equation}
This value already deviates about $13 \%$ from the zero-th order approximation and already agrees very well with exact numerical result.

\subsection{Photon current}
In Fig. 2(c) in the main text we plot the profile of the mean  photon current $\vec j_p(\vec r_i)$. On the discrete lattice we define the $x$ ($y$) component of $\vec j_p(\vec r_i)$  as the average between the number of photons per unit of time passing from site $\vec r_i$ to site $\vec r_i+ \vec e_x$ ($\vec r_i+ \vec e_y$) and the number of photons per unit of time passing from site  $\vec r_i- \vec e_x$ ($\vec r_i- \vec e_y$) to site $\vec r_i$. 
Explicitly, the two components of the photon current are defined as 
\begin{eqnarray}
\vec j^{x}_p(\vec r_i)&=&i \frac{J}{2}\left[ \left(e^{i\phi_x} \Psi^{\dag}(\vec{r}_i+\vec e_{x}) - e^{-i\phi_x} \Psi^{\dag}(\vec{r}_i-\vec e_{x}) \right)\Psi(\vec{r}_i) - {\rm H. c. }\right],\\
\vec j^{y}_p(\vec r_i)&=&i \frac{J}{2}\left[ \left(e^{i\phi_y} \Psi^{\dag}(\vec{r}_i+\vec e_{y}) - e^{-i\phi_y} \Psi^{\dag}(\vec{r}_i-\vec e_{y}) \right)\Psi(\vec{r}_i) - {\rm H. c. }\right].
\end{eqnarray}
The plots in Fig. 2(c) in the main text show a vector plot of the expectation value of this operator with respect to the exact single-excitation wavefunction $|\psi\rangle(t_\pi)$. 

To connect this expression to the usual current density operator in the continuum limit we identify $\vec j_c(\vec r_i)= \vec j_p(\vec r_i)/l_0$ and $\Psi_c(\vec r_i)= \Psi(\vec r_i) /l_0$, such that $[\Psi_c(\vec r),\Psi^\dag_c(\vec r')]\approx \delta (\vec r-\vec r')$ in the limit $l_0\rightarrow0$.  Then, by expanding $\vec j_p(\vec r_i)$  to lowest order in $l_0$ we obtain
\begin{equation}
\vec j_c(\vec r)=\frac{1}{2m} \left[\Psi_c^{\dag}(\vec{r})\left(-i\hbar \vec \nabla \right)\Psi_c(\vec{r}) - {\rm H.c.} \right]  - \frac{e}{m} \vec A(\vec r) \Psi_c^{\dag}(\vec{r})\Psi_c(\vec{r}) .
\end{equation}




\section{Photon propagator and Landau Green's function}
Since the photons are noninteracting, the dynamics of the photonic lattice can be fully captured by the single-photon Green's function,
\begin{equation}
G(t ,\vec r_i, \vec r_j)= \langle {\rm vac}| \Psi(\vec r_i,t )\Psi^\dag(\vec r_j,0)|{\rm vac}\rangle=\sum_\lambda f_\lambda(\vec r_i) f^*_\lambda(\vec r_j) e^{-i\omega_\lambda t}.
\end{equation}
In the long-wavelength limit and for moderate magnetic fields, the mode functions $f_\lambda(\vec r_i)$ can be approximated by Landau orbitals and
\begin{equation}
G(t, \vec{r}_i, \vec{r}_j) \simeq \sum_{\ell k} \Phi_{\ell k}(\vec{r}_i) \Phi^*_{\ell k}(\vec{r}_j)  e^{-i\omega_\ell t}.
\end{equation}
Note that for a simple square lattice it is in principle still possible to obtain an exact expression for $G(t, \vec{r}_i, \vec{r}_j)$ in terms of a continued fraction \cite{Ueta1997S}. However, this expression must still be evaluated numerically and does not offer much physical insight in the considered regime of moderate field strengths, where the continuum approximation is more intuitive and provides sufficiently accurate results.

To carry out the sum over the index $k$ in the continuum limit, it is convenient to re-express the Landau orbitals as
\begin{equation}
\Phi_{\ell k}(\vec r) = \sqrt{\alpha}\braket{k | \mathcal{D}(\xi ) | \ell },
\end{equation}
where $\mathcal{D}(\xi )=e^{\xi a^\dag -\xi^* a}$ is the displacement operator for a bosonic mode with annihilation operator $a$ and $|\ell,k\rangle$ are the corresponding number states \cite{galuber1969S}. This identification allows us to make use of the general relation for displacement operators, $\mathcal{D}^\dag(\xi) \mathcal{D}(\beta) = \mathcal{D}(\beta-\xi) e^{-\frac{1}{2}(\xi \beta^* - \xi^* \beta)}$, to show that 
\begin{equation}
\begin{split}
\sum_{k} \Phi_{\ell k}(\vec r_i) \Phi^*_{\ell k}(\vec r_j)   & = \alpha \sum_k \braket{\ell | \mathcal{D}^\dag( \xi_j) | k }\braket{ k | \mathcal{D}  (\xi_i) | \ell  }
\\
& = \alpha \braket{\ell  |  \mathcal{D}^\dag( \xi_j) \mathcal{D}  (\xi_i) | \ell }=  \alpha e^{\frac{1}{2}(\xi_i \xi_j^*- \xi_i^*\xi_j)} \braket{\ell | \mathcal{D} (\xi_i-\xi_j) | \ell },\\
& = \sqrt{\alpha} e^{i \theta_{ij} } \Phi_{\ell \ell} (\vec r_i-\vec r_j),
\end{split}
\end{equation}
where $\theta_{ij}=-i(\xi_i\xi_j^*-\xi_i^*\xi_j)/2=-(x_iy_j-x_jy_i )/(2l_B^2)$. 
Note that by going from the first to the second line we have used the completeness relation, $\mathbbm{1} \simeq  \sum_k |k \rangle \langle k |$. This assumes that the degeneracy of each Landau level is sufficiently large, which corresponds to having a system sufficiently larger than the magnetic length $l_B$ not to feel finite-size effects.
Under these approximations the total lattice Green's function reduces to the continuum Green's function of a single charged particle \cite{Ueta1992S}. It can be explicitly expressed as a sum over all Landau levels
\begin{equation}\label{eq:SuppG}
G(t, \vec{r}_i, \vec{r}_j)\simeq  \sum_{\ell } G_\ell(\vec{r}_i, \vec{r}_j) e^{-i\omega_\ell t},
\end{equation}
where 
\begin{equation}
G_\ell(\vec{r}_i, \vec{r}_j) = \sqrt{\alpha} e^{i\theta_{ij}} \Phi_{\ell \ell} (\vec r_i-\vec r_j).
\end{equation}
Remarkably, resuming the degeneracy of each Landau level, the only non-vanishing contributions to the Green's function comes from the orbitals $\Phi_{\ell \ell}$ with zero angular momentum, $L_z \sim k-\ell = 0$.

\subsection{Gauge transformations}
The vector potential $\vec{A}$ is only defined up to the gradient of an arbitrary function. Once a representation of the vector potential is fixed, one can still change to an equivalent representation by adding the gradient of a suitable function, $\vec{A}(\vec{r}) \longmapsto \vec{A}(\vec{r}) - \vec \nabla \Lambda(\vec{r})$. In order to have a gauge independent Schr\"odinger equation (and thus, consistently, gauge independent observables) the phase of the wave function must change accordingly,  $\psi \longmapsto e^{ie\Lambda/\hbar}\psi$. The same it is true for the photonic Green's function which transforms under gauge transformations as
\begin{equation}
G (\tau, \vec{r}_i, \vec{r}_j) \longmapsto e^{ie(\Lambda(\vec{r}_i) - \Lambda(\vec{r}_j))/\hbar}G (\tau, \vec{r}_i, \vec{r}_j).
\end{equation}
The immediate consequence of this is that the Green's function must split in two parts, a gauge invariant amplitude, and a gauge dependent phase, where the amplitude depends only on the distance $|\vec{r}_i - \vec{r}_j|$:
\begin{equation}
G (\tau, \vec{r}_i, \vec{r}_j) = e^{ i\theta_{ij}}G^{\rm inv.} (\tau, |\vec{r}_i - \vec{r}_j| ).
\end{equation}
In the intermediate flux regime, where the continuum approximation holds, $G^{\rm inv} (\tau, |\vec{r}_i - \vec{r}_j| ) \sim \sum_{\ell} \Phi_{\ell \ell} (|\vec r_i-\vec r_j|)e^{-i\omega_\ell t}$, while $\theta_{ij}$ is still depends on the choice of the gauge.

\subsection{Landau-level projector}
Equation \eqref{eq:SuppG} shows that in the continuum limit the photonic's Green's function can be written as the sum over the components  $G_{\ell}(\vec{r}_i, \vec{r}_j)$ for each band. This decomposition is particularly relevant when the splitting $\omega_c$ is sufficiently large and emitters couple dominantly to a single band. The $G_{\ell}(\vec{r}_i, \vec{r}_j)$ are real-space representations of the band-projector operators $\hat{\mathcal{P}}_{\ell}$ \cite{Alicki1993S,Assaad1995S}, i.e., 
\begin{equation}
 \langle r_i |\hat{\mathcal{P}}_{\ell} | r_j \rangle=G_{\ell}(\vec{r}_i, \vec{r}_j) = \sum_k \Phi_{\ell k} (\vec{r}_i ) \Phi_{\ell k}^* (\vec{r}_j ).
 \end{equation}
In this sense, one can define photonic operators $\tilde \Psi_{\ell}(\vec r_i) = \sum_j \, G_{\ell}(\vec{r}_i, \vec{r}_j) \Psi(\vec{r}_j)$, which are  field operators projected onto a single Landau level. In general, these operators are not orthogonal and therefore the bosonic operators $B_{\ell n}$ introduced in Eq. (8) in the main text are linear combinations of those projected operators. By evaluating the commutators
\begin{equation}
\begin{split}
[B_{\ell n},B_{\ell n'}^\dag] & = \sum_{m,m'} K^{-1}_{n m} (K^{-1}_{n' m'})^{*} \sum_{ij} G_{\ell } (\vec r_e^{\, m}, \vec{r}_i)  G^*_{\ell } (\vec r_e^{\,m'}, \vec{r}_j) \delta_{ij}
\\
& = \sum_{m,m'} K^{-1}_{n m} (K^{-1}_{n' m'})^{*} G_{\ell } (\vec r_e^{\,m}, \vec r_e^{\,m'}) 
\\
& =  \left[K^{-1} G (K^{-1})^\dag\right]_{nn'} \overset{!}{=} \delta_{n n'}
\end{split}
\end{equation}
we see that the operators $B_{\ell n}$ represent an independent set of modes when $K K^{\dag} = G$, where $G$ is an $N\times N$ matrix with elements $G_{\ell } (\vec r_e^{\,m}, \vec r_e^{\,m'})  $.  For explicit calculations we diagonalize $G$ and take the square root of each of its eigenvalues $\chi_i$. After transforming back to the original basis we obtain 
\begin{equation}
K = U^{\dag} {\rm diag}(\sqrt{\chi_1}, \sqrt{\chi_2} \cdots \sqrt{\chi_{N}})U,
\end{equation}
where $U$ is the diagonalizing matrix. Note that the matrix-$K$ is not uniquely defined and here we always use the positive square roots of the $\chi_i$.
In the case of $N=2$ emitters we obtain the result
\begin{equation}
K = \frac{1}{\sqrt{ {\rm Tr}[G] + 2 \sqrt{ {\rm det}[G] }}} \left( G + \sqrt{{\rm det}[G]} \mathbbm{1} \right),
\end{equation}
or, explicitly,  
\begin{equation}
K = \sqrt{\frac{\alpha}{2}}
\begin{pmatrix}
\sqrt{1 + \sqrt{1-e^{-|\xi_0|^2}L_{\ell}^2(|\xi_0|^2) } } & \frac{e^{-|\xi_0|^2/2}L_{\ell}(|\xi_0|^2)}{\sqrt{1 + \sqrt{1-e^{-|\xi_0|^2}L_{\ell}^2(|\xi_0|^2) } }} \\
\frac{e^{-|\xi_0|^2/2}L_{\ell}(|\xi_0|^2)}{\sqrt{1 + \sqrt{1-e^{-|\xi_0|^2}L_{\ell}^2(|\xi_0|^2) } }} & \sqrt{1 + \sqrt{1-e^{-|\xi_0|^2}L_{\ell}^2(|\xi_0|^2) } }
\end{pmatrix},
\end{equation}
where $\xi_0 = |\vec{r}_1 - \vec{r}_2|/\sqrt{2 l_{B}^2}$.

\section{Resonant interactions in the single excitation sector}
We consider the dynamics in the single excitation sector, meaning that we restrict the dynamics to states of the form
\begin{equation}
|\psi\rangle(t)= \left[ \sum_{n=1}^N c_n(t)\sigma_+^n + \sum_\lambda \varphi_\lambda(t) \Psi_\lambda^\dag\right]|g\rangle|{\rm vac}\rangle,
\end{equation}
where $\lambda$ labels the single photon eigenstates.
Plugging this ansatz into the time dependent Schr\"odinger equation, $i\partial_t | \psi \rangle = H | \psi \rangle$, where $H$ is given in Eq. (2)
in the main text, we obtain the following equations of motion
\begin{equation}\label{eq:single_ex_one_atom_dyn}
\begin{split}
i \dot{c}_n & = \left(\omega_e - i\gamma_e/2\right) c_n + g\sum_{\lambda} f_{\lambda}(\vec{r}_e^{\, n}) \varphi_{\lambda},
\\
i \dot{\varphi}_{\lambda} & = (\omega_\lambda - i \gamma_p/2) \varphi_{\lambda} + g \sum_m f^*_{\lambda}(\vec{r}_e^{\, m})c_m,
\end{split}
\end{equation}
where we included a decay of the emitters with rate $\gamma_e$ and photon losses with rate $\gamma_p$.
We can formally integrate the second equation (for the photon populations) and obtain
\begin{equation}\label{eq:photon_wave_fun_modespace}
\varphi_{\lambda}(t) = - i g\sum_m f^*_{\lambda}(\vec{r}_e^{\, m})\int_0^t e^{-i(\omega_\lambda-i\gamma_p/2)(t-t')}c_m(t') dt',
\end{equation}
where we assumed $\varphi_{\lambda}(t=0) = 0$ (i.e., initially there are no photons in the system). 
By reinserting this result into the equations for the emitter's amplitude we end up with 
\begin{equation}\label{eq:many_atoms_sing_ex_integral_eq}
\dot{c}_n(t) = - i(\omega_e-i\gamma_e/2) c_n - g^2 \sum_m \int_0^t G(t-t',\vec{r}_e^{\, n}, \vec{r}_e^{\, m})e^{-\gamma_p(t-t')/2} c_m(t') dt'.
\end{equation} 
This result is still completely general and used to produce the numerical results presented in Fig. 2 in the main text.

\subsection{Spontaneous emission in a non-magnetic lattice}
We consider here in detail the single emitter case. Considering the transformation $c_e(t) \mapsto c_e(t) e^{- i(\omega_e-i\gamma_e/2)t}$, Eq. \eqref{eq:many_atoms_sing_ex_integral_eq} can be rewritten as
\begin{equation}
\dot{c}_e(t) = - g^2\int_0^t K(t-t') e^{\bar \gamma (t-t')/2} c_e(t') dt',
\end{equation} 
where $\bar \gamma = \gamma_e - \gamma_p$ and the integral kernel is given by
\begin{equation}
K(t) = \int_{-\infty}^{+\infty} \rho(\vec{r}_e, \omega) e^{-i(\omega-\omega_e)t } d\omega,
\end{equation}
with $\rho(\vec{r}_e, \omega)  = \sum_\lambda |f_{\lambda}(\vec{r}_e)|^2\delta(\omega-\omega_\lambda)$, as defined in the main text.
In an infinitely large system, the density of states becomes a smooth function of $\omega$. When the coupling is small and the emitter's resonance is sufficiently far away from eventual singular points \cite{Gonzalez-Tudela2017S}, we can approximate it as a constant, $\rho(\vec{r}_e, \omega) \simeq \rho(\vec{r}_e, \omega_e)=\tau/(2\pi)$. 
In this way the integral kernel can be approximated by a delta function, $K(t-t') \simeq \tau \delta(t-t')$, which is evaluated at the upper bound of the integral. We then recover the usual exponential decay 
\begin{equation}
\dot{c}_e(t) = - \frac{g^2 \tau}{2} c_e(t).
\end{equation}

In a 2D system with eigenmodes $f_{\lambda} \sim e^{i \vec{k}\cdot \vec{r}}$ and an approximately quadratic dispersion, $\omega_k \simeq \omega_b +J |\vec k|^2$, we obtain $\tau \simeq 1/(2J)$ and 
\begin{equation}
\Gamma \simeq \frac{g^2}{2J}.
\end{equation}
For smaller lattices, delimited by sharp edges, the emitted photons will be reflected at the boundaries and for longer times the decay of the emitter will deviate from a purely exponential shape. To avoid such boundary effects we have included in the numerical simulations in Fig. 2(a) in the main text a larger photon loss rates at the edges to mimic an infinitely extended system. To implement the dissipative boundaries it is more convenient to rewrite Eq. \eqref{eq:single_ex_one_atom_dyn} using the photon's wave function $\varphi(t, \vec{r}) = \sum_{\lambda} f_{\lambda}(\vec{r}) \varphi_{\lambda}(t)$, which gives (in general for $N$ emitters)
\begin{equation}
\begin{split}
i \dot{c}_n & = \left(\omega_e - i\gamma_e/2\right) c_n + g\varphi(t, \vec{r}_e^{\, n}),
\\
i \dot{\varphi}(t, \vec{r}_i) & = \sum_j\left[- J_{ij} + (\omega_p - i \tilde{\gamma}_p(\vec{r}_i)/2 )\delta_{ij} \right] \varphi(t, \vec{r}_j) + g \sum_m \delta_{m i} c_m ,
\end{split}
\end{equation}
where now we introduced a space dependent photonic dissipation $\tilde{\gamma}_p(\vec{r})$. In our simulations we used a Fermi-function-like profile
\begin{equation}
\tilde{\gamma}_p(\vec{r}) = \gamma_p + \frac{\gamma_{\rm edge}}{1+\exp[-(r-R_0)/2]}.
\end{equation}
Typically we tune the parameters such as $R_0 \simeq L/2$, where $L$ is the characteristic size of the system, and $\gamma_{\rm edge}\simeq \gamma_p \times 10^{3}$. Note that these additional loss channels do not affect the evolution of the coupled emitter-photon state in the case of a finite $\alpha$.


\subsection{Flat-band approximation}
When the light-matter coupling $g$ is larger than the width of the $\ell$-th band, but still much smaller than the gap to the other bands, we can make a resonance approximation. To do so we discard the contributions from all the other bands and treat the $\ell$-th band as degenerate. Under these assumptions, i.e, $|\omega_e- \omega_{\ell k}| \ll g $ and  $g \ll |\omega_{\ell k} - \omega_{\ell\pm 1 k'}|$, and by changing into a damped rotating frame, $c_n(t) \mapsto c_n(t) e^{-i(\omega_e - i\gamma_p/2 ) t}$, we obtain the approximate result
\begin{equation}
\dot{c}_n(t) \simeq   - \frac{\bar \gamma}{2} c_n - g^2\sum_m \int_0^t G_\ell (\vec{r}_e^{\, n}, \vec{r}_e^{\, m}) c_m(t') dt',
\end{equation}
where $\bar \gamma = \gamma_e - \gamma_p$ is the difference between the loss rates. Taking the time derivative of this equation we obtain a set of second order differential equations for $N$  coupled harmonic oscillators,
\begin{equation}\label{eq:atom_atom_res_eff_int_eq}
\ddot{c}_n(t) = -\frac{\bar \gamma}{2} \dot{c}_n(t) - g^2 \sum_m G_\ell (\vec{r}_e^{\, n}, \vec{r}_e^{\, m}) c_m(t).
\end{equation}

\subsection{LPP spectrum} 
By taking the Fourier transform of the $c_n(t)$ in Eq.~\eqref{eq:atom_atom_res_eff_int_eq} we obtain the eigenvalue equation
\begin{equation}\label{eq:fourier_atom_atom_res_eff_int_eq}
(\omega^2 + i\omega \bar \gamma /2 - \Omega^2) c_n(\omega) =  g^2 \sum_{m\neq n} G_\ell (\vec{r}_e^{\, n}, \vec{r}_e^{\, m}) c_m(\omega),
\end{equation}
from which we can derive the complex eigenvalues of the resonant LPPs, which represent the resonance frequencies and the decay rates of the coupled eigenmodes.  After transforming back into the original frame, these complex eigenvalues are
\begin{equation}\label{eq:generic_singl_ex_eigenvalues}
\omega_{\nu} = \omega_e - i\frac{\gamma_e + \gamma_p}{4} \pm \Omega\sqrt{1+ \Lambda_{\nu}-\bar \gamma^2/(16\Omega^2) },
\end{equation}
where the $\Lambda_{\nu}$ are the eigenvalues of the  matrix
\begin{equation}
\mathcal{M} = \frac{1}{\alpha}
\begin{pmatrix}
0 & G_\ell(\vec{r}_e^{\, 1}, \vec{r}_e^{\, 2}) & G_\ell(\vec{r}_e^{\, 1}, \vec{r}_e^{\, 3}) & \cdots & G_\ell(\vec{r}_e^{\, 1}, \vec{r}_e^{\, N}) \\
G_\ell(\vec{r}_e^{\, 2}, \vec{r}_e^{\, 1}) & 0 & G_\ell(\vec{r}_e^{\, 2}, \vec{r}_e^{\, 3}) & \cdots & G_\ell(\vec{r}_e^{\, 2}, \vec{r}_e^{\, N}) \\
\vdots & & \ddots & &\cdots \\
\end{pmatrix}.
\end{equation}
The symmetry between the different single-excitation states in respectively the upper and lower polaritons that is visible in Eq.~\eqref{eq:generic_singl_ex_eigenvalues} as a $\pm$ in front of the $\Lambda_\nu$-dependent square root and in the single-excitation manifold of Fig~3(a)
as a symmetry of the branches with respect to $\omega_e$ is a consequence of the fact that the emitters are resonantly coupled to a single and degenerate Landau level. Quite interestingly, a similar symmetry with respect to $n\omega_e$ is also visible in the $n$-excitation manifold, e.g. in the $n=2$ manifold of Fig~3(a)

For the example of three equidistant emitters,
\begin{equation}
\frac{1}{\alpha} G_\ell(\vec{r}_e^{\, n}, \vec{r}_e^{\, m}) = e^{-\frac{d^2 }{4l_B^2}} L_\ell^0 \left( \frac{d^2}{2l_B^2}\right) e^{i\theta_{nm}}
\end{equation}
and $\Lambda_\nu =e^{-d^2/(4l_B^2)} L_\ell^0 \left( d^2/(2l_B^2)\right) \times \lambda_\nu$, where $\lambda_\nu$ are the eigenvalues of the reduced matrix 
\begin{equation}
\tilde{\mathcal{M}} =\begin{pmatrix}
0 & e^{i \theta_{12}} & e^{-i \theta_{31}} \\
e^{-i \theta_{12}} & 0 & e^{i \theta_{23}} \\
e^{i \theta_{31}} & e^{-i \theta_{23}} & 0 \\
\end{pmatrix}.
\end{equation}
Therefore, the $\lambda_\nu$ are determined by the solutions of
\begin{equation}
\lambda^3 -3\lambda-2\cos(\theta_\triangle)=0,
\end{equation}
which only depend on the gauge invariant sum of all the phases, 
\begin{equation}
\theta_\triangle=\theta_{12}+\theta_{23}+\theta_{31}= \frac{A_\triangle}{l_B^2}= \frac{e B A_\triangle}{\hbar}.
\end{equation}
The solutions are explicitly given by
\begin{equation}
\lambda_\nu = 2 \cos\left( \frac{\theta_\triangle+2\pi \nu}{3}\right).
\end{equation}


\section{Band-gap chiral excitation flow}

The condition of perfect chiral or non-chiral excitation flow in an equilateral triangle of emitters, strongly detuned from any specific Landau level, is related to the eigenvalues of $\tilde{J}_{nm}$. In particular, a fully chiral or completely non-chiral flow appears, when one of the single excitation eigenvalues become zero or when two of them become degenerate. 
Indeed the single excitation sector of the equilateral triangular system is fully described just looking at the eigenvalues/eigenstates of the band-gap interaction itself
\begin{equation}\label{eq:chiral_dyn_matrix_3atoms}
\tilde{J} = G_0
\begin{pmatrix}
0 & e^{i \theta_{12}} & e^{i \theta_{13}} \\
e^{-i \theta_{12}} & 0 & e^{i \theta_{23}} \\
e^{-i \theta_{13}} & e^{-i \theta_{23}} & 0 \\
\end{pmatrix},
\end{equation}
where $G_0 = g^2/(\omega_a-\omega_\ell) \Phi_{\ell\ell}(|\vec r_a^{\,n}-\vec r_a^{\,m}|)$ can be regarded just as a constant, since we consider an equilateral triangle geometry.
The characteristic polynomial of the system is given by
\begin{equation}
\lambda^3 - 3 G_0^2 \lambda - 2G_0^3 \cos( \theta_{\Delta}) = 0,
\end{equation}
which is exactly the same polynomials used to find the eigenvalues in the resonant case (up to a scale factor $G_0$). We have that perfect chirality/non-chirality are realised, respectively, when $\theta_{\Delta} = n\pi/2$  with $n$ odd-integer, or  $\theta_{\Delta}= n\pi$  with $n$ even-integer.
This information is just given by the determinant of the effective interaction, which is ${\rm det}[\tilde{J}_{nm}] = 2G_0^3 \cos( \theta_{\Delta}) $. When ${\rm det}[\tilde{J}_{nm}] = 0$ we have perfect chirality, on contrary, when ${\rm det}[\tilde{J}_{nm}] = \pm 2G_0^3$ chirality is lost, as the magnetic field were turned off.
This can be worked out exactly, by considering that $c_n(t) = \sum_{\nu} \sum_m c_m(t=0)f_{\nu} (m) f_{\nu}(n) e^{-i\lambda_{\nu} t}$, where $c_n(t)$, for $n=1,2,3$ is the population of the $n$-th emitter, and $f_{\nu}(n)$, $\lambda_{\nu}$ are, respectively the eigenvectors, eigenvalues of the dynamical matrix \eqref{eq:chiral_dyn_matrix_3atoms}. Assuming the excitation is initially loaded just in the first emitter, i.e. $c_n(t=0) = \delta_{0n}$, and considering $ \theta_{\Delta} = n\pi/2$  we have
\begin{equation}
\begin{split}
|c_1(t)| & = \bigg| \frac{1}{3} + \frac{2}{3}\cos\left[\sqrt{3}G_0 t \right] \bigg|
\\
|c_2(t)| & = \bigg| \frac{1}{3} + \frac{2}{3}\cos\left[\sqrt{3}G_0 t + \frac{4\pi}{3} \right] \bigg|
\\
|c_3(t)| & = \bigg| \frac{1}{3} + \frac{2}{3}\cos\left[\sqrt{3}G_0 t + \frac{2\pi}{3} \right] \bigg|
\end{split}
\end{equation}
This solution clearly shows that the chirality emerges from the $2\pi/3$ phase shift between the three different populations oscillations.

\section{Disorder}\label{sec:SuppDisorder}
\begin{figure}
\centering
	\includegraphics[width=\columnwidth]{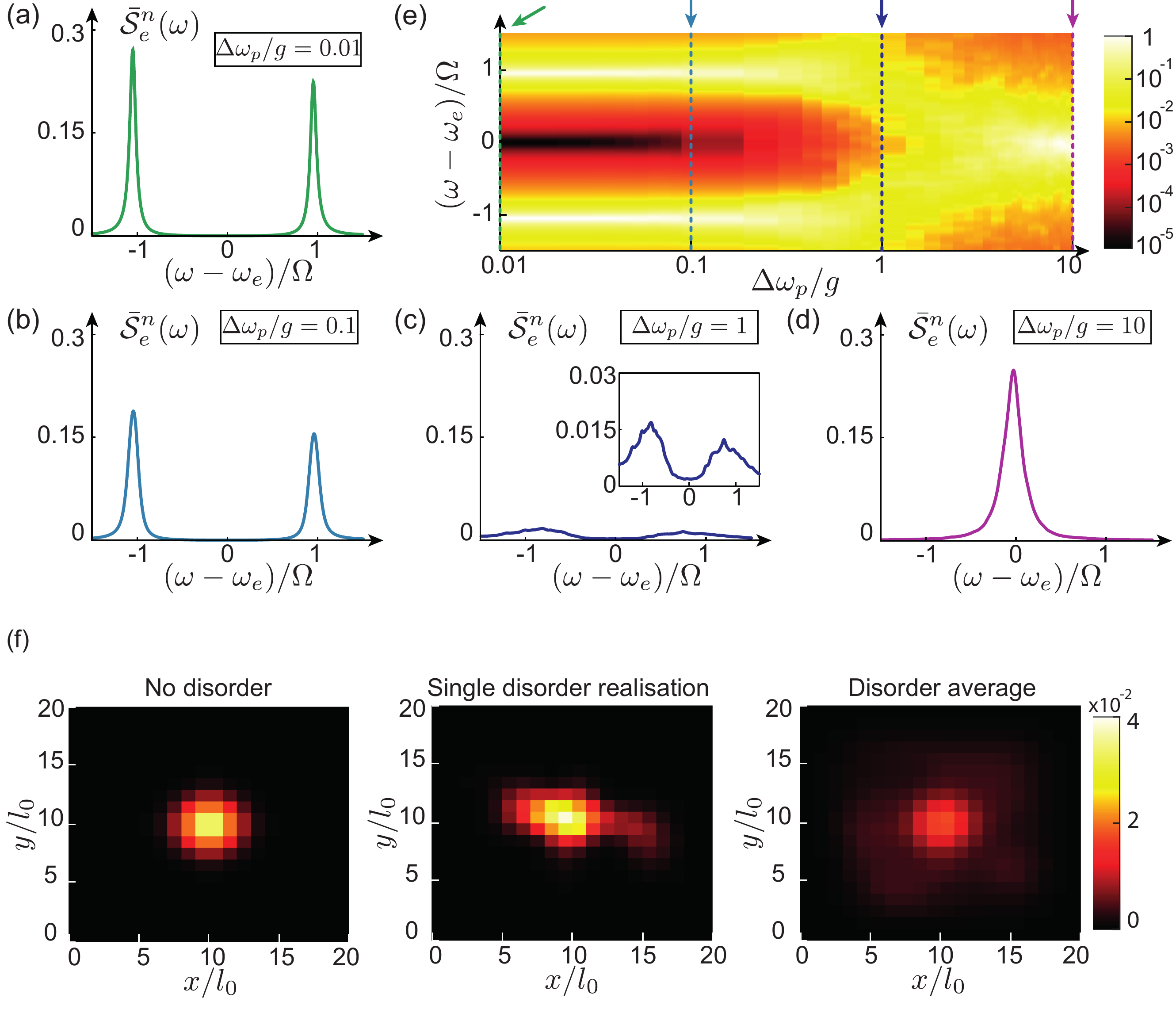}
	\caption{(a-d) Disorder averaged excitation spectrum $\bar{\mathcal{S}}_e^n(\omega)$ for fixed value of the disorder strength (as indicate in each plot). Each plot is averaged over $N_{\rm dis}=1000$ realisations. (e) Disorder averaged excitation spectrum $\bar{\mathcal{S}}_e^n(\omega)$ as a function of  $\Delta \omega_p$. For each value of $\Delta \omega_p$ the excitation spectrum is averaged over $N_{\rm dis}=50$ realisations.  (f) Plot of the photon wavefunction $|\varphi(\vec{r}_n)|^2$ of the lowest LPP. The disorder strength in this plot is chosen as $\Delta \omega_p/g = 0.7$. The left panel shows the case without disorder, the center panel the wavefunction for a single disorder realisation and the right panel depicts the average over $N_{\rm dis} = 200$ realisations.
For all figures we have assumed a $M=20\times 20$ photonic lattice, $\alpha=0.08$, $\delta_e/J = 0.47$ (corresponding to the resonance with the $\ell=0$ Landau level) and $g/J=0.08$. }
	\label{Fig:Supp_disorder}
\end{figure}

All our calculations in the main text are based on the assumption of an ideal lattice for the photons. In practice, fabrication uncertainties will result, for example, in random local offsets of the bare photon frequency $\omega_p$, which will affect the energies and wavefunctions of the photons. To estimate the effect of disorder on the LPPs, 
we now replace $\omega_p$ with a random offset at every site, $\tilde{\omega}_p^i = \omega_p + \delta \omega_p^i$, where $\delta \omega_p^i$ is sampled from a Gaussian distribution centered around zero and with a width $\Delta \omega_p$.  

In the case of emitters resonantly coupled to the lattice, we expect the main physics is barely affected by the disorder, provided that $\omega_c \gg g \gg \Delta \omega_p$ (where for higher Landau levels $\omega_c$ is replaced by the frequency difference between two neighbouring  levels).
We now illustrate this point more explicitly on the simplest case of the single emitter. We consider the excitation spectrum, as defined in the main text,
\begin{equation}
\mathcal{S}^{n}_e(\omega) = 
\left| \braket{G | \sigma_-^n  \frac{1}{H - \omega  - i\frac{\gamma_e}{2}\sum_{m}\sigma_+^m\sigma_-^m }\sigma_+^n | G} \right|^2,
\end{equation}
where $H$ is now affected by the onsite disorder, as defined above. 
A good quantity that will provide a clear visualization of the effect of disorder is the average excitation spectrum defined as
\begin{equation}
\bar{\mathcal{S}}_e^n(\omega) = \frac{1}{N_{\rm dis}} \sum_{k=1}^{N_{\rm dis} }\mathcal{S}_e^n(\omega),
\end{equation}
where $N_{\rm dis}$ is the number of disorder realizations. In each realization the onsite energies $\tilde{\omega}_p^i$ for each site are chosen randomly, as described above.
In Fig. \ref{Fig:Supp_disorder}(a-e) we plot the resulting average excitation spectrum for a single emitter, in resonance with the lowest Landau level. 
This plot shows that the Rabi splitting (and thus the presence of the chiral bound state) is almost unaffected for disorder strengths up to $\Delta \omega_p/g\lesssim 1$ and even up to values of $\Delta \omega_p/g\simeq 1$ the splitting is still visible. In this regime the main effect of disorder is a broadening of the lines. Only at larger disorder strengths the LPPs break up and the excitation spectrum reduces to a single line centered around the emitter frequency. Note that in the considered regime of interest, $\omega_c> g$, the condition $\Delta \omega_p< g$ also implies that the disorder does not mix the Landau levels. Therefore, the chiral properties of the LPPs remain preserved. 
Moreover in Fig. \ref{Fig:Supp_disorder}(f) we visualize the spatial profile of the photon wavefunction, $|\varphi(\vec{r}_n)|^2$, of the lowest LPP in the presence of disorder. Note that for this example a rather larger disorder strength of $\Delta \omega_p/g = 0.7$ has been assumed. Even in this regime the Landau level is on average still clearly recognizable (right panel), although in a single disorder realization its rotational symmetry is already partially lost (center panel).

%

\begin{figure}
\centering
	\includegraphics[width=\columnwidth]{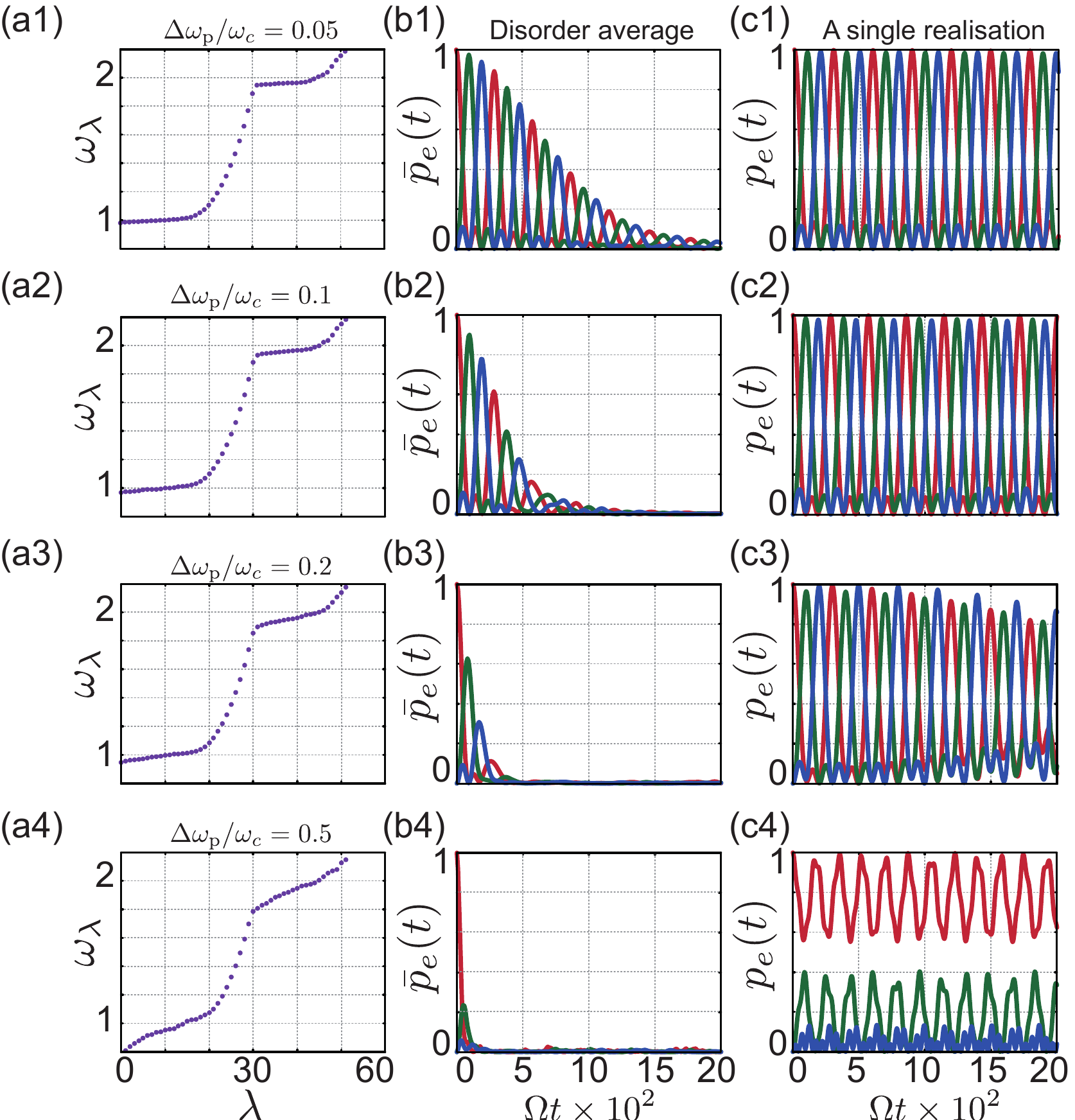}
	\caption{(a1-a4) Plot of the lowest eigenvalues $\omega_\lambda$ of $H_{\rm ph}$  in the presence of disorder and for a $31\times 31$ triangular lattice. (b1-b4) Disorder averaged evolution of the excited state population, $\bar{p}_e$, for three equidistant emitters with $d/l_0=4$. Here the bar denotes the average over $N_{\rm dis}=100$ realizations, see Eq.~\eqref{eq:pebar}. The disorder strength used is the same as reported in the respective panel (a) plots. (c1-c4) Single realization of the population's time evolution under the same conditions as the panel (a-b) plots. The other parameters used in all the plots are $\omega_{\rm p} =9.5$, $\omega_e=0.5$, $J=0.75$, $g=0.1$, $\alpha = 1/(16\sqrt{3})$, $d/l_0=4$, $\gamma=10^{-5}$ (all frequencies are given in arbitrary units). }
	\label{Fig:dis_bandgap}
\end{figure}

For emitters that are detuned from the nearest Landau level we expect that the constraint on the tolerable level of disorder can be further relaxed and the sufficient condition to observe all non-resonant effects detailed in the main text is to have ${\rm min}\{\omega_c, |\omega_e-\omega_\ell|\} \gg \Delta \omega_p$. 
Large quantitative and qualitative deviations from the main results of this work are expected once the disorder approaches the scale of the cyclotron frequency, affecting both the amplitude, but also the phase of the emerging dipole-dipole interactions. 
In Fig. \ref{Fig:dis_bandgap} we report the result of another numerical experiment. In each row of this figure labels (1), (2), (3) and (4) correspond to the disorder strengths $\Delta \omega_{\rm p} = 0.05, 0.1, 0.2, 0.5$. In the column (a1-a4) we report the lowest part of the spectrum of $H_{\rm ph}$ for different disorder strengths. The column (b1-b4) shows the time evolution of the average population
\begin{equation}\label{eq:pebar}
\bar{p}_e(t) = \bigg| \frac{1}{N_{\rm dis.}} \sum_{\rm dis.} c_e(t) \bigg|^2,
\end{equation}
where $N_{\rm dis.}$ is the number of disorder realisations considered (in our simulations $N_{\rm dis.} = 100$), and $\sum_{\rm dis.}$ is the sum over all these realisations. The average time evolution defined in this way allows us to study dephasing due to the disorder. 

When the disorder strength is much smaller than the cyclotron frequency it just weakly affects the effective magnetic phase between the emitters. This can be seen in Fig. \ref{Fig:dis_bandgap}(b1-b2), where up to $\Delta \omega_{\rm p}/\omega_c \lesssim 0.1 $ the average time evolution still exhibits many oscillations before it is washed out by dephasing. Looking at single disorder realizations, Fig. \ref{Fig:dis_bandgap}(c1-c2), it is likely to see perfect chirality (up to very long times) with just small ripples in the dynamics.
When the disorder starts to be comparable to the cyclotron frequency the magnetic phases in the effective model $H_{\rm eff}$ starts to be affected.  This is shown in Fig. \ref{Fig:dis_bandgap}(b3-c3) for $\Delta \omega_{\rm p}/\omega_c = 0.2$. While for a single realization the emitters still undergoes many chiral oscillations, the average dynamics is already strongly damped. This means that the emitter dynamics is still governed by the effective Hamiltonian $H_{\rm eff}$, but the tunneling amplitudes and phases are no longer predictable.
In the last example, we set $\Delta \omega_{\rm p}/\omega_c \leq 0.5$. The Landau gap is still open, but the width of each level is now comparable to the gap between them. In this regime an approximate description in terms of separated Landau levels is no longer possible. This can be seen immediately from the time evolution of the average evolution, Fig. \ref{Fig:dis_bandgap}(b4), where the dephasing is so strong that no single oscillation can occur. Also for most individual disorder realizations the chiral flow of excitations is broken, as shown in Fig. \ref{Fig:dis_bandgap}(c4).

\section{Experimental implementations}
To probe the main properties of LPPs in experiments, we need a photonic lattice with a synthetic magnetic field, two-level emitters coupled to the photonic lattice with a strength $g$ that exceeds the bare loss rates $\gamma_p$ and $\gamma_e$, and a sufficiently low level of disorder in the lattice such that the formation of discrete Landau levels and a hybridization between photons and emitters is still possible, see Sec.~\ref{sec:SuppDisorder}. While in the long term these conditions might be achievable in different platforms in the optical and microwave regime, we here briefly discuss a setup proposed in Ref. \cite{Anderson2016S} and implemented in Ref. \cite{Owens2018S}, where the physics of LPPs can already be probed with existing technology. Ref. \cite{Anderson2016S} describes a 2D magnetic photonic lattice composed of 3D microwave resonators, which contain a magnetic material. By applying an external field, this magnetic component breaks the time reversal symmetry of the local modes and allows one to realize in a scalable way large lattices with effective fields of $\alpha=1/4$ or $\alpha=1/6$ \cite{Anderson2016S}. The on-site frequency of these microwave resonator can take values in the range of hundreds of MHz to tens of GHz, with quality factors $Q \sim 10^3-10^5$. The tunneling rate between neighboring resonators can be engineered to be $J \sim 100$ MHz. 
At the same time, the onsite disorder, $\Delta \omega_{p}$, for such microwave resonators can be quite small and in similar coupled resonator arrays values of $\Delta \omega_{p} \lesssim 1$ MHz have been reported \cite{Mirhosseini2018S,Saxberg2019S}.  The two-level systems can be represented in this system by superconducting qubits, where qubit-resonator couplings in the range of $g \sim 1-100$ MHz are readily achievable (see, e.g., Refs.~\cite{Mirhosseini2018S,Saxberg2019S,Kim2020S}). Superconducting qubits are typically designed with a frequency of about $\omega_e\sim 3-5$ GHz and exhibit coherence times of about $0.1-1$ ms, which translates into typical decay rates of about $\gamma_e \sim$ kHz. Since the qubit frequency is easily tunable by several 100 MHz, it can be easily made resonant with one of the Landau levels. 

Based on these estimates, we can readily identify a set of experimentally realistic parameters, which are close to the values assumed for most of the results discussed in the main text:
\begin{table}[h]
	\centering
	\begin{tabular}{|c|c|c|c|c|c|c|}
		\hline 
		$\omega_{\rm p}$ & $\omega_e$  & $J$  & $g$ & $\gamma_{p,e}$ & $\,\,\Delta \omega_p\,\,$ & \\  \hline 
		$\,\,5.4 \times 10^3\,\,$ & $\,\,5\times 10^3\,\,$  & $\,\,100\,\,$  & $\,\,20\,\,$  & $\,\,0.05\,\,$ & $\,\,1\,\,$ & \,\,MHz\,\,\\
		\hline
	\end{tabular}
\end{table}

Note that while in the setting described in Ref. \cite{Anderson2016S} the magnetic flux assumes the fixed values $\alpha =1/4,1/6$, these values are still in the intermediate regime, where a continuum approximation is valid. To achieve arbitrary values of $\alpha$ more flexible approaches, such as demonstrated in Ref. \cite{Roushan2017S} can be used, where arbitrary phase patterns can be imprinted by external driving fields.

\end{document}